\documentclass[showpacs,preprintnumbers,amsmath,amssymb,eqsecnum]{revtex4}

\usepackage{graphicx}
\usepackage{color} 
\usepackage{amsmath}%
\usepackage{amsfonts}%
\usepackage{amssymb}%
\usepackage{hyperref}

\begin{document}

\newcommand{\pderiv}[2]{\frac{\partial #1}{\partial #2}}
\newcommand{\deriv}[2]{\frac{d #1}{d #2}}
\newcommand{\eq}[1]{Eq.~(\ref{#1})}  
\newcommand{\infint}{\int \limits_{-\infty}^{\infty}}

\title{Thermodynamic Framework for Compact q-Gaussian Distributions}

\vskip \baselineskip

\author{Andre M.~C. Souza$^{1,4}$}
\thanks{E-mail address: amcsouza@ufs.br}

\author{Roberto F.~S. Andrade$^{2,4}$}
\thanks{E-mail address: randrade@ufba.br}

\author{Fernando D. Nobre$^{3,4}$}
\thanks{Corresponding author: E-mail address: fdnobre@cbpf.br}

\author{Evaldo M.~F. Curado$^{3,4}$}
\thanks{E-mail address: evaldo@cbpf.br}

\address{
$^{1}$ Departamento de F\'\i sica, Universidade Federal de Sergipe
\\49100-000 \hspace{5mm} S\~ao Crist\'ov\~ao - SE \hspace{5mm} Brazil \\
$^{2}$ Instituto de F\'\i sica, Universidade Federal da Bahia
\\40210-340 \hspace{5mm} Salvador - BA \hspace{5mm} Brazil \\
$^{3}$Centro Brasileiro de Pesquisas F\'{\i}sicas, Rua Xavier Sigaud 150 \\
22290-180 \hspace{5mm} Rio de Janeiro - RJ  \hspace{5mm} Brazil \\
$^{4}$National Institute of Science and Technology for Complex Systems \\
Rua Xavier Sigaud 150 \\
22290-180 \hspace{5mm} Rio de Janeiro - RJ  \hspace{5mm} Brazil}

\date{\today}

\begin{abstract}
Recent works have associated systems of particles, characterized
by short-range repulsive interactions and evolving under overdamped motion,
to a nonlinear Fokker-Planck
equation within the class of nonextensive statistical
mechanics, with a nonlinear diffusion contribution
whose exponent is given by $\nu=2-q$.
The particular case $\nu=2$ applies to
interacting vortices in type-II
superconductors, whereas $\nu>2$ covers systems of particles
characterized by short-range power-law interactions, where correlations
among particles are taken into account.
In the former case, several studies
presented a consistent thermodynamic
framework based on the definition of an effective temperature
$\theta$ (presenting experimental values much higher than
typical room temperatures $T$, so that thermal noise could be neglected),
conjugated to a generalized entropy $s_{\nu}$ (with $\nu=2$).
Herein, the whole thermodynamic scheme is revisited and extended to systems
of particles interacting repulsively, through short-ranged potentials,
described by an entropy $s_{\nu}$, with $\nu>1$,
covering the $\nu=2$ (vortices in type-II
superconductors) and $\nu>2$ (short-range power-law interactions)
physical examples.
One basic requirement concerns
a cutoff in the equilibrium distribution $P_{\rm eq}(x)$,
approached due to a confining external harmonic
potential, $\phi(x)=\alpha x^2/2$ ($\alpha>0$).
The main results achieved are:
(a) The definition of an effective temperature $\theta$ conjugated
to the entropy $s_{\nu}$;
(b) The construction of a Carnot cycle,
whose efficiency is shown to be $\eta=1-(\theta_2/\theta_1)$,
where $\theta_1$ and $\theta_2$ are the effective temperatures
associated with two isothermal transformations,
with $\theta_1>\theta_2$;
(c) Thermodynamic potentials, Maxwell relations, and response
functions.
The present thermodynamic framework,
for a system of interacting particles under the above-mentioned
conditions, and associated to an entropy $s_{\nu}$, with $\nu>1$,
certainly enlarges the possibility of experimental verifications.

\vskip \baselineskip




\end{abstract}
\maketitle

\section{Introduction}


An appropriate thermodynamic description for a system of interacting
particles represents a fundamental intent in physics and may
become a great challenge in many cases~\cite{reif,balian,reichl}.
Depending on the particular type of interactions, the outstanding
framework of statistical mechanics may turn intractable, at least from the
analytical point of view, in such a way that numerical approaches
emerged as important tools in the recent years.
The elegant connection between the microscopic world (described by
statistical mechanics) and the macroscopic one (described
by thermodynamics) occurs usually through the entropy concept,
so that the knowledge of the entropy associated to a given
system becomes a crucial step in this pathway. Although
the standard procedure should be applied for systems at
equilibrium, the connection of the entropy concept with dynamics
represents a remarkable result of nonequilibrium statistical
mechanics~\cite{balian,reichl,balakrishnan}.
The statistical entropy $s$ is defined as a functional depending only
on the probabilities of a physical system~\cite{balian},
i.e., $s \equiv s\left\{P_{i}(t) \right\}$, for a discrete
set of states, where $P_{i}(t)$ represents the probability for finding
the system in a state $i$, at time $t$, while
$s \equiv s[P(x,t)]$, for continuous states, where
$x$ usually denotes the position in a one-dimensional space.
The association with dynamics may occur by means
of the following procedures:
(i) The statistical entropy may be extremized under certain constraints,
in order to yield an equilibrium probability that
coincides with the stationary-state distribution obtained from
some equations describing the time evolution of the probabilities
(e.g., a Fokker-Planck equation~\cite{risken});
(ii) An H-theorem, which may be proven
by considering the statistical entropy and a given equation for the
time evolution of the probabilities~\cite{balian,reichl,balakrishnan,risken}.

These connections among entropies and dynamics were extended
for generalized entropic forms, mostly by making use of nonlinear
Fokker-Planck equations (NLFPEs)~\cite{frankbook}; a typical NLFPE,
relevant for the present work, is given by

\begin{eqnarray}
\label{eq:qtsallisnlfpetemp}
\mu \ {\partial P(x,t) \over \partial t} & = &
-{\partial [A(x)P(x,t)] \over \partial x}
+\nu D \ {\partial \over \partial x}
\left\{[\lambda P(x,t)]^{\nu -1}{\partial P(x,t) \over \partial x}\right\}
\nonumber \\ &  &
+k T \ {\partial^{2} P(x,t) \over \partial x^{2}}~,
\end{eqnarray}

\vskip \baselineskip
\noindent
where $\mu$ stands for a friction coefficient, $\nu$ a real number, and $\lambda$
represents a characteristic length of the system. One should notice
that, due to the normalization condition, the probability $P(x,t)$ presents
dimension $[{\rm length}]^{-1}$ and consequently, the characteristic length
$\lambda$ was introduced in the nonlinear diffusion term for dimensional reasons.
Moreover, the diffusion constant
$D$ may result from a coarse-graining procedure (see, e.g.,
Refs.~\cite{zapperiprl2001,andradeprl2010,ribeiropre2012}), being directly
related to particle-particle interactions, thus depending on
the physical system under investigation.
The above equation also takes into account the effects of a
heat-bath at a temperature $T$ (with $k$ standing for
the Boltzmann constant), whose contribution may be obtained
in the standard way, through the
introduction of a thermal noise in the system~\cite{reichl}.
Additionally, $A(x)=-d\phi(x)/dx$ corresponds to an external force
derived from a confining potential $\phi(x)$, being fundamental for
the approach to equilibrium, as well as for the resulting
form of the probability distribution in the long-time limit.
The particular case $\nu=2$ has been much explored recently,
being shown to be directly related to a system of interacting
vortices, relevant for type-II
superconductors~\cite{zapperiprl2001,andradeprl2010,ribeiropre2012,%
ribeiroepjb2012,ribeiropre2014,nobrepre2012,curadopre2014,%
randradeepl2014,nobrepre2015,ribeiropre2015,ribeiropre2016};
in this application, $\lambda$ represents
the London penetration length, and
it was shown that $D \gg kT$, so that thermal
effects could be neglected~\cite{nobrepre2012}.
Herein, we will be concerned with physical systems for which this
approximation applies, leading to,

\begin{equation}
\label{eq:qtsallisnlfpe}
\mu \ {\partial P(x,t) \over \partial t} = -{\partial [A(x)P(x,t)] \over \partial x}
+\nu D \ {\partial \over \partial x}
\left\{[\lambda P(x,t)]^{\nu -1}{\partial P(x,t) \over \partial x}\right\}~,
\end{equation}

\vskip \baselineskip
\noindent
corresponding to the NLFPE introduced in
Refs.~\cite{plastinophysa1995,tsallispre1996}, being
associated with Tsallis entropy~\cite{tsallis88}, through
the identification $\nu=2-q$, where $q$ represents the
well-known entropic index. The above equation became an
useful tool for dealing with a wide range of natural
phenomena, like those related to anomalous
diffusion~\cite{tsallisbook,tsallisreview2011,tsallisreview2014}.

Generalized forms of the H-theorem, making use of NLFPEs,
were developed by many authors in the recent
years~\cite{shiino01,kaniadakispla2001,frankdaff01,%
frank2002,shiino2003,chavanis03,schwaemmle07a,schwaemmle07b,%
schwaemmle09,chavanis2008,ribeiro11}; particularly, the NLFPE
in~\eq{eq:qtsallisnlfpe} was shown to be associated to Tsallis'
entropy,

\begin{equation}
\label{eq:tsallisentnu}
s_{\nu}[P] = {k  \over \nu-1} \ \left\{ 1- \lambda^{\nu-1}
\int_{-\infty}^{\infty}dx \ [P(x,t)]^{\nu}
\right\}~,
\end{equation}

\vskip \baselineskip
\noindent
through a well-defined sign for the time derivative of the free-energy
functional, i.e., $(df/dt) \leq 0$, where

\begin{equation}
\label{eq:freeenergy}
f[P]= u[P] - \theta s_{\nu}[P]~;
\qquad u[P] =  \int_{-\infty}^{\infty} dx \ \phi(x) P(x,t)~.
\end{equation}

\vskip \baselineskip
\noindent
It is important to remind that the characteristic length
$\lambda$ appearing in~\eq{eq:tsallisentnu} follows from an H-theorem
[proven by making use of~\eq{eq:qtsallisnlfpe}] and it leads to the correct dimension
for the entropy.
Moreover, the time-dependent solution of~\eq{eq:qtsallisnlfpe},
for an initial condition $P(x,0)=\delta(x)$ and a harmonic
external force, $A(x)=-\alpha x$ $(\alpha>0)$,
was found as~\cite{plastinophysa1995,tsallispre1996},

\begin{equation}
\label{eq:tsallisdist}
P(x,t)=B(t) [ 1 - b(t)(\nu-1)x^{2} ]_{+}^{1/(\nu-1)}~,
\end{equation}

\vskip \baselineskip
\noindent
where $[y]_{+}=y$, for $y>0$, zero otherwise, and the time-dependent coefficients
$B(t)$ and $b(t)$ are related to each other in order to preserve the
normalization of $P(x,t)$ for all times $t$. The above solution,
usually called $q$-Gaussian (also considering the idenfication $\nu=2-q$),
coincides, in the long-time limit, with the equilibrium distribution that comes
out from the extremization of the entropy in~\eq{eq:tsallisentnu} by
imposing usual constraints, such as probability normalization and
the internal energy definition of~\eq{eq:freeenergy}.

The purpose of this work is to present a consistent
thermodynamic framework, associated with the equilibrium
distribution of~\eq{eq:tsallisdist}, for a certain range of values
of the real parameter $\nu$.
The work is motivated by recent results
of Ref.~\cite{vieirapre2016}, where a coarse-graining
approach was developed for an interacting
system, and correlations were taken into account, leading to the NLFPE
of~\eq{eq:qtsallisnlfpe} with $q<0$, i.e., $\nu>2$.
Although we are not aware of similar approaches valid
for systems of interacting particles in wider intervals for the parameter $\nu$,
the thermodynamic framework to be presented herein applies to $\nu>1$;
this will certainly cover many phenomena within the scope of anomalous
diffusion, associated
with~\eq{eq:qtsallisnlfpe}~\cite{tsallisbook,tsallisreview2011,tsallisreview2014}.
The procedure is based on the definition of an effective
temperature $\theta$, conjugated to the entropic form
of~\eq{eq:tsallisentnu}, typical of nonextensive statistical mechanics.
The present approach extends previous
ones, carried for the particular case $\nu=2$, and
expected to hold for a system of interacting particles,
whose final mechanical equilibrium is reached after a long-time
evolution, under overdamped
motion~\cite{nobrepre2012,curadopre2014,%
randradeepl2014,nobrepre2015,ribeiropre2015,ribeiropre2016}.
Like in the previous works, the following framework holds
for typical effective temperatures $\theta \gg T$, so that
the usual thermal effects (associated with the temperature $T$)
can be neglected.

In the next section we describe
the coarse-graining procedure developed in Ref.~\cite{vieirapre2016},
applied to the corresponding equations of motion, in such a way
to associate a physical system
to the NLFPE of~\eq{eq:qtsallisnlfpe}.
The correlations are taken into account in a qualitative way, and
we discuss the conditions under which the approach should hold.
In Section III we present a definition for an effective temperature $\theta$,
introduce the equilibrium distribution $P_{\rm eq}(x)$, which corresponds
essentially to a $q$-Gaussian distribution with $q<1$, being
characterized by a cutoff $x_{e}$.
As a consequence, although the initial motivation concerns a
physical system considered in Ref.~\cite{vieirapre2016},
which was shown to be associated with
a NLFPE with $q<0$, the present thermodynamic framework
applies to a wider range of values of $q$, holding
for any $q<1$, i.e., $\nu>1$.
Furthermore, in Section III we calculate
the internal energy and entropy in terms of $\theta$.
In Section IV we define the thermal contact between two systems, obtain
relations involving the corresponding initial effective temperatures and
final temperature, and an equivalent formulation for the Zeroth Law of
thermodynamics.
In Section V we propose an infinitesimal form for the Fist Law, define
transformations and explore the Carnot cycle, obtaining the celebrated
expression for its efficiency.
In Section VI we consider Legendre transformations in order to
derive thermodynamic potentials, Maxwell relations, and
obtain well-known structures for the thermodynamic potentials.
Finally, in Section VII we present our main conclusions.

\section{Illustration of a Physical System}

Next, we describe the coarse-graining approximation introduced
in Ref.~\cite{vieirapre2016}, leading to the NLFPE
of~\eq{eq:qtsallisnlfpe}; this procedure was carried for
a system of $N$ interacting particles, under overdamped motion,
so that the equation of motion of a particle $i$, in
a medium with an effective
friction coefficient $\mu$, may be written as

\begin{equation}
\label{eq:mov}
\mu \mathbf{v}_i=\mathbf{F}_i^{\rm int}+\mathbf{F}_i^{\rm ext}  \quad
\quad (i=1,2, \cdots , N),
\end{equation}

\vskip \baselineskip
\noindent
where $\mathbf{v}_i$ stands for its velocity, and the
condition of  overdamped motion neglects the inertial
contribution $m(d\mathbf{v}_i/dt)$.
Moreover, the terms on the right-hand-side
depict the forces acting on the particle, due to internal and
external agents, respectively.
The main novelty of the approach of Ref.~\cite{vieirapre2016},
with respect to previous
ones~\cite{zapperiprl2001,andradeprl2010,ribeiropre2012,%
ribeiroepjb2012,ribeiropre2014}, consists
in taking into account the role played by
particle-particle correlations in the contribution
$\mathbf{F}_i^{\rm int}$, as will be shown later. As usual, one has

\begin{equation}
\label{eq:ppforces}
\mathbf{F}_i^{\rm int} = {1 \over 2} \sum_{j \neq i}
{B}_{ij} (r_{ij}) \ \hat{\mathbf{r}}_{ij}~,
\end{equation}

\vskip \baselineskip
\noindent
where the factor $1/2$ comes to compensate the
double counting of interactions, the distance between
particles $i$ and $j$ is
$r_{ij}=|\mathbf{r}_i - \mathbf{r}_j|$, and
$\hat{\mathbf{r}}_{ij} = (\mathbf{r}_i - \mathbf{r}_j)/r_{ij}$
is a vector defined along the axis of each pair of
particles. Moreover, ${B}_{ij} (r_{ij})$ represents the functional form
of the particle-particle interactions,
which varies according to the physical system.

Hence, for a description in terms of continuous variables,
one performs a coarse
graining by introducing a local density of particles at time $t$,
$\rho({\mathbf{r}},t)$, and based on the conservation
of the total number of particles, one can make use of the continuity equation,

\begin{equation}
\label{eq:conteq}
\frac{\partial \rho({\mathbf{r}},t)}{\partial t} = \nabla\cdot \mathbf{J}~,
\end{equation}

\vskip \baselineskip
\noindent
where the current density
$\mathbf{J}=\rho\mathbf{v}$ was defined. Supposing
that the local density varies smoothly around a certain point
$\tilde{\mathbf{r}}$, one may expand
$\rho({\mathbf{r}},t) \approx \rho(\tilde{\mathbf{r}},t) +
{\mathbf{r}} \cdot \nabla \rho({\mathbf{r}},t)|_{\mathbf{r}={\tilde{\mathbf{r}}}}$,
so that the effect of the remaining $(N-1)$ particles on a given
particle may be written in terms of the force $\mathbf{F}^{\rm int}$,
expressed as an integral in a $d$-dimensional space,

\begin{equation}
\label{eq:cgrainppforces}
\mathbf{F}^{\rm int} = {1 \over 2} \int d^{d}r\rho(\mathbf{r},t) {B} (r) \hat{\mathbf{r}}
\approx {1 \over 2} \int d^{d}r  [{\mathbf{r}} \cdot \nabla
\rho({\mathbf{r}},t)]_{\mathbf{r}={\tilde{\mathbf{r}}}} {B}(r) \hat{\mathbf{r}}~,
\end{equation}

\vskip \baselineskip
\noindent
where we have used the result that, due to symmetry, the integral
over $\rho(\tilde{\mathbf{r}},t)$ vanishes identically.
Without loss of generality, we take
$\nabla \rho(\tilde{\mathbf{r}},t)$ along the $\hat{\mathbf{r}}$
direction, so that performing the integral over angles
one obtains the constant factor $\Omega_{d}$,

\begin{equation}
\label{eq:ppforcesaconst}
\mathbf{F}^{\rm int} \approx {\Omega_{d} \over 2}
\nabla \rho(\tilde{\mathbf{r}},t) \int dr \ r^{d} {B}(r)~,
\end{equation}

\vskip \baselineskip
\noindent
and we suppose that the local density changes slowly
within the range of interactions. The above equation may
still be written in the form

\begin{equation}
\label{eq:constadef}
\mathbf{F}^{\rm int} \approx a\nabla \rho(\tilde{\mathbf{r}},t)~;
\quad a = {\Omega_{d} \over 2} \int_{0}^{\infty} dr \ r^{d} {B}(r)~,
\end{equation}

\vskip \baselineskip
\noindent
where we are assuming short-range particle-particle interactions;
indeed, in the case $d=2$ ($\Omega_{2}=2\pi)$, \eq{eq:constadef}
recovers the result of
Refs.~\cite{zapperiprl2001,andradeprl2010,ribeiropre2012,%
ribeiroepjb2012,ribeiropre2014}.

Then, redefining $\tilde{\mathbf{r}} \rightarrow {\mathbf{r}}$,
and using the external force as
$\mathbf{F}^{\rm ext} = -A(x) \hat{\mathbf{x}}$, where $A(x)$ is associated
with some type of confining potential in the $\hat{\mathbf{x}}$ direction,
one chooses this direction to be special for attaining the final equilibrium.
Hence, considering fixed values for all coordinates, except $x$,
substituting these results into~\eq{eq:conteq} and using~\eq{eq:mov}, one
gets that

\begin{equation}
\label{eq:rhodiffusioneq}
\mu \frac{\partial \rho({\mathbf{r}},t)}{\partial t} = \nabla \cdot
\left\{ \rho({\mathbf{r}},t)
\left[ a  \nabla \rho({\mathbf{r}},t) + \mathbf{F}^{\rm ext}
\right] \right\}
= \frac{\partial}{\partial x} \left\{ \rho({\mathbf{r}},t)
\left[ a \ \frac{\partial \rho({\mathbf{r}},t)}{\partial x}
- A(x) \right] \right\}~,
\end{equation}

\vskip \baselineskip
\noindent
which may be easily written in the form of the NLFPE
of~\eq{eq:qtsallisnlfpe} for $\nu=2$; in this case,
the quantity $a$ of~\eq{eq:constadef} will be
directly related to $D$ and will depend
on the characteristic parameters of the particle-particle
interactions (see, e.g., Ref.~\cite{ribeiropre2012}).

The contribution of Ref.~\cite{vieirapre2016} consisted in
assuming that correlations between particles produce
an exclusion region of finite radius around each particle,
$r_{0} \equiv r_{0}(\rho)$, which becomes smaller as the concentration
$\rho$ increases. In this way, the lower limit in the integral
of~\eq{eq:constadef} was set to $r_{0}(\rho)$, leading to

\begin{equation}
\label{eq:ppforcesarho}
\mathbf{F}^{\rm int} \approx  a(\rho)  \nabla \rho({\mathbf{r}},t)~; \quad
a(\rho) = {\Omega_{d} \over 2} \int_{r_{0}(\rho)}^{\infty} dr \ r^{d} {B}(r)~,
\end{equation}

\vskip \baselineskip
\noindent
so that~\eq{eq:rhodiffusioneq} becomes

\begin{equation}
\label{eq:arhodiffusioneq}
\mu \frac{\partial \rho({\mathbf{r}},t)}{\partial t}
= \frac{\partial}{\partial x} \left\{ \rho({\mathbf{r}},t)
\left[ a(\rho) \ \frac{\partial \rho({\mathbf{r}},t)}{\partial x}
- A(x) \right] \right\}~,
\end{equation}

\vskip \baselineskip
\noindent
resulting in the relevant modification $a \rightarrow a(\rho)$.
Although the modification introduced in~\eq{eq:ppforcesarho} appears as
a qualitative correction in the
resultant force $\mathbf{F}^{\rm int}$, it takes into account
correlations, which may become relevant in some cases, particularly
for short distances between particles.

Considering particle-particle interactions associated to short-range
repulsive potentials of the type
$V(r)= f_{0} \lambda (r/\lambda)^{-\xi}$ ($\xi > d$),
where $\lambda$ represents a characteristic length of the system,
whereas $f_{0}$ is a positive constant, the
authors of Ref.~\cite{vieirapre2016} have shown that
$r_{0}(\rho) \propto \rho^{-1/d}$ leads to
the NLFPE of~\eq{eq:qtsallisnlfpe}, with $q=1-(\xi/d) < 0$,
i.e., $\nu>2$.
In the cases where the potential $V(r)$ leads to a finite integral
in~\eq{eq:constadef}, one results with a constant value for $a$;
therefore, this particular choice for $r_{0}(\rho)$ yields,
for high enough concentrations, $r_{0} \rightarrow 0$,
so that $\lim_{\rho \rightarrow \infty} a(\rho) = a$.
As an  example of this limit, one has the system of type-II
superconducting vortices interacting repulsively
in two dimensions (corresponding to $\nu=2$), for which
$a=2\pi f_{0}\lambda^{3}$~\cite{zapperiprl2001,andradeprl2010,ribeiropre2012,%
ribeiroepjb2012,ribeiropre2014,nobrepre2012,curadopre2014,%
randradeepl2014,nobrepre2015,ribeiropre2015,ribeiropre2016}.
The good agreement between numerical data from molecular-dynamics
simulations and the analytical result of~\eq{eq:tsallisdist} for the
equilibrium distribution,
in both cases $d=2$ and $d=3$, and several choices for $\xi$~\cite{vieirapre2016},
supports the assumptions considered above and more specifically,
the lower limit introduced in~\eq{eq:ppforcesarho}.

Therefore, restricting to $d=2$ for simplicity, the particle-particle
interactions become
$B(r)=-[dV(r)/dr]= \xi f_{0} (r/\lambda)^{-\xi-1}$ ($\xi > 2$). Now, we consider
$r_{0}(\rho) = z \rho^{-1/2}$, where $z$ depends on $\xi$ in such a way that

\begin{equation}
\label{eq:zxiomega}
\lim_{\xi \rightarrow 2^{+}} \left( \frac{z^{2-\xi}}{\xi-2} \right) = \omega~,
\end{equation}

\vskip \baselineskip
\noindent
with $\omega$ being a positive finite number; consequently,
one obtains from~\eq{eq:ppforcesarho},

\begin{equation}
\label{eq:arho}
a(\rho) = \pi \int_{z \rho^{-1/2}}^{\infty} dr \ r^{2} {B}(r) =
{\pi \xi z^{2-\xi} f_{0} \lambda^{\xi+1} \over \xi-2} \rho^{-1+\xi/2}~,
\end{equation}

\vskip \baselineskip
\noindent
so that the condition of~\eq{eq:zxiomega} leads to
$\lim_{\xi \rightarrow 2^{+}}a(\rho)=2\omega \pi f_{0} \lambda^{3}$.
One should notice that considering the above form for $a(\rho)$ and
defining the probability for finding a given particle with a coordinate $x$ at time $t$
as $P(x,t)=(L_{y}/N)\rho(x,t)$, one obtains
the NLFPE of~\eq{eq:qtsallisnlfpe} by setting
$\xi=2(\nu-1)=2(1-q)$.
In what follows we develop a thermodynamic framework, based on
the NLFPE of~\eq{eq:qtsallisnlfpe}, its associated entropy
[\eq{eq:tsallisentnu}], and the corresponding equilibrium
distribution [\eq{eq:tsallisdist}].

\section{Effective-Temperature and Equilibrium Distribution}

Before starting with the basic definitions of thermodynamic quantities,
it is important to remind some conditions, and define some
nomenclature to be used, as described below.
\noindent
(i) Although one may deal with a general dimensionality
$d$, herein, for simplicity, we will restrict ourselves
to $d=2$, so that the system
of particles is confined in a two-dimensional rectangular
box of sizes $L_{x}$ and
$L_{y}$ (measured in units of the basic length $\lambda$).
\noindent
(ii) In order to conform with the solution of~\eq{eq:tsallisdist}, the
whole procedure will be developed by considering
a harmonic external force, $A(x)=-\alpha x$ $(\alpha>0)$.
\noindent
(iii) Due to the H theorem, the stationary-state solution of~\eq{eq:qtsallisnlfpe},
which coincides with the distribution that maximizes the
entropy of~\eq{eq:tsallisentnu},
will be referred from now on as equilibrium distribution; it corresponds to
the limit $t \rightarrow \infty$
of the distribution in~\eq{eq:tsallisdist}.
\noindent
(iv) Despite of the fact that the physical system described in the previous
section is characterized by $q<0$, the whole scheme that follows holds
in general for $q<1$. In such cases, the distribution
of~\eq{eq:tsallisdist} presents a compact support, so that the integration
limits in Eqs.~(\ref{eq:tsallisentnu}) and~(\ref{eq:freeenergy})
should be replaced by finite values, $ \pm {\bar x}(t)$,
with the cutoff of the equilibrium distribution being given by
$x_{e}=\lim_{t \rightarrow \infty} {\bar x}(t)$~\cite{nobrepre2012,%
curadopre2014,randradeepl2014,nobrepre2015,ribeiropre2015,ribeiropre2016}.

The effective temperature comes directly from the coarse-graining
procedure leading to the NLFPE of~\eq{eq:qtsallisnlfpe}; in the case of
the physical application described above, it can be defined as

\begin{equation}
\label{eq:gentempappl}
k\theta \equiv D = {\pi z^{2(2-\nu)}(\nu-1) f_{0} \lambda \over \nu(\nu-2)}
\left( {N \lambda \over  L_{y}} \right)^{\nu-1}~,
\end{equation}

\vskip \baselineskip
\noindent
where the identification $(\xi/2)=\nu-1$ was done, yielding the
condition $\nu>2$.
However, although the illustration of the previous
section is characterized by $q<0$, we shall present herein a procedure that
applies in general for $q<1$.
For that, we will extend the definition above
by considering~\eq{eq:zxiomega}, which becomes

\begin{equation}
\label{eq:znuomega}
\lim_{\nu \rightarrow 2^{+}} \left[ \frac{z^{2(2-\nu)}}{2(\nu-2)} \right] = \omega~.
\end{equation}

\vskip \baselineskip
\noindent
Since the motivation of the present approach is to extend previous
works, carried for the particular case $\nu=2$, from now on we set
$\omega=1$, in such a way to recover from~\eq{eq:arho},
$\lim_{\xi \rightarrow 2^{+}}a(\rho)=2\pi f_{0} \lambda^{3}$, obtained
in Ref.~\cite{ribeiropre2012} for the system of type-II superconducting vortices.
In this way, all previous results for the thermodynamics of the equilibrium state
of this system~\cite{nobrepre2012,curadopre2014,%
randradeepl2014,nobrepre2015,ribeiropre2015,ribeiropre2016} may be
recovered in this particular limit.
Therefore, we write the effective temperature of~\eq{eq:gentempappl}
as

\begin{equation}
\label{eq:gentemp}
k\theta \equiv D = {b_{\nu} f_{0} \lambda \over \nu}
\left( {N \lambda \over  L_{y}} \right)^{\nu-1}~,
\end{equation}

\vskip \baselineskip
\noindent
where

\begin{equation}
\label{eq:bnu}
b_{\nu} = {\pi (\nu-1) \over \nu-2} \ z^{2(2-\nu)}~; \qquad (\nu \geq 2)~,
\end{equation}

\vskip \baselineskip
\noindent
leading to $b_{2}=2\pi$.
Moreover, the precise expression of the coefficient $b_{\nu}$ for the interval
$1 < \nu < 2$ will depend on an associated physical system,
for which the present approach may be applicable, although it should be a positive
real number and expected to match with the above result by means of
$\lim_{\nu \rightarrow 2^{-}}b_{\nu}=2\pi$.
However, one should call the attention to the important
property (valid for any $\nu>1$)
that the effective temperature $\theta$ depends directly
on the linear density of particles $(N/L_{y})$, which
may (hopefully) be a controllable quantity in many physical systems.
As an example, for the system of interacting vortices,
relevant for type-II superconductors~\cite{poole},
recent advances in experimental techniques have made
the density of vortices a controllable
quantity~\cite{lee,derenyi,villegas,zhu}, leading
to the desirable possibility of varying $\theta$.
Moreover, in this particular application, the parameters
of~\eq{eq:gentemp} yield typical effective temperatures
$\theta$ much higher than room temperature $T$, i.e.,
$\theta \gg T$~\cite{nobrepre2012}, so that thermal effects were
neglected in the analyses of Refs.~\cite{nobrepre2012,%
curadopre2014,randradeepl2014,nobrepre2015,ribeiropre2015,ribeiropre2016}.
Herein, we will restrict ourselves to physical systems for which the
parameters of~\eq{eq:gentemp} also lead to $\theta \gg T$.
The appropriateness of the above definition for an effective temperature,
as well as of the conditions imposed herein,
will be shown throughout this paper.

The equilibrium solution corresponds to the limit $t \rightarrow \infty$
of the distribution in~\eq{eq:tsallisdist}~\cite{andradeprl2010,ribeiropre2012},
being herein expressed as

\begin{equation}
\label{eq:equilpx}
P_{\rm eq}(x) = \frac{\Gamma(\frac{3}{2}+\frac{1}{\nu-1} )}{\sqrt{\pi} \
\Gamma(\frac{\nu}{\nu-1} )(x_{e})^{\frac{\nu+1}{\nu-1}}} \
\left( x_{e}^2 - x^2 \right)^{1/(\nu-1)}
=  \frac{\Gamma(\frac{3}{2}+\frac{1}{\nu-1} )}{\sqrt{\pi} \
\Gamma(\frac{\nu}{\nu-1} ) \ x_{e}} \
\left[ 1 - \left({x \over x_{e}} \right)^{2} \right]^{1/(\nu-1)}~,
\end{equation}

\vskip \baselineskip
\noindent
with $|x| < x_{e}$, where

\begin{equation}
\label{eq:xe}
x_{e} = C_{\nu} \lambda \left( \frac{k\theta}{\alpha \lambda^{2}}
\right)^{1/(\nu+1)}~.
\end{equation}

\vskip \baselineskip
\noindent
In the equation above $C_{\nu}$ is a dimensionless $\nu$-dependent coefficient,

\begin{equation}
C_{\nu} = \left\{ \frac{2 \nu}{\nu-1} \right\}^{1/(\nu+1)}
\left\{ \frac{(\nu+1)\Gamma(\frac{1}{2}+\frac{1}{\nu-1}
)}{2(\nu-1)\sqrt{\pi} \ \Gamma(1+\frac{1}{\nu-1} )}
\right\}^{(\nu-1)/(\nu+1)}~,
\end{equation}

\vskip \baselineskip
\noindent
which, in the particular case $\nu=2$, investigated in
Refs.~\cite{randradeepl2014,curadopre2014,nobrepre2015}, becomes
$C_{2}=3^{1/3}$. From~\eq{eq:xe} one notices
the physically important result that $x_{e}$ gets larger
by increasing the effective temperature $\theta$, for any $\nu>1$.
From the equilibrium distribution, one can calculate the internal energy
of~\eq{eq:freeenergy},

\begin{equation}
\label{eq:intennu}
u = \int_{-x_{e}}^{x_{e}}dx \ {\alpha x^{2}
\over 2} \  P_{\rm eq}(x) =  {\nu-1 \over 3\nu-1} {\alpha x_{e}^{2}
\over 2} = {\nu-1 \over 6\nu-2} C_{\nu}^{2} \ \alpha \lambda^2 \left(
\frac{k\theta }{\alpha\lambda^2} \right)^{2/(\nu+1)},
\end{equation}

\vskip \baselineskip
\noindent
which is directly related to the variance,

\begin{equation}
\label{eq:averx2}
\langle x^{2} \rangle = \int_{-x_{e}}^{x_{e}}dx \ x^{2} \ P_{\rm eq}(x)
= {\nu-1 \over 3\nu-1} C_{\nu}^{2} \ \left(
\frac{k\theta }{\alpha\lambda^2} \right)^{2/(\nu+1)}\lambda^2 ,
\end{equation}

\vskip \baselineskip
\noindent
showing that the equilibrium distribution spreads for increasing values
of the effective temperature.

Therefore, the effective temperature $\theta$ is related to the variance in
particle positions, following $\theta \propto \langle x^2 \rangle^{(\nu+1)/2}$.
This behavior is a signature of the NLFPE of~\eq{eq:qtsallisnlfpe}, and
it should be contrasted with the standard classical dilute
gas~\cite{reif,balian,reichl}, for which the
temperature $T$ is related linearly to the second moment of the
corresponding velocity
probability distribution, i.e., $T \propto \langle v^{2} \rangle$.
Since we are dealing with a system of particles for which
the thermal noise was neglected ($T/\theta \simeq 0$), the final
state is in fact a mechanical equilibrium, where its
effective temperature $\theta$ comes out to be related to the
variance $ \langle x^2 \rangle$ in a nonlinear way.

Similarly to the quantities above, one can calculate the entropy
of~\eq{eq:tsallisentnu},

\begin{equation}
\label{eq:enttheta}
s_{\nu} = \frac{k}{\nu-1}
\left\{ 1- B_{\nu} [C_{\nu}]^{\frac{3\nu-1}{\nu-1}} \left(
\frac{\alpha \lambda^{2} }{k\theta} \right)^{\frac{\nu-1}{\nu+1}}
\right\}~,
\end{equation}

\vskip \baselineskip
\noindent
where we have introduced another dimensionless $\nu$-dependent coefficient,

\begin{equation}
B_{\nu} = \frac{ \sqrt{\pi}\; \Gamma(1+\frac{\nu}{\nu-1})}{
\Gamma(\frac{3}{2}+\frac{\nu}{\nu-1})} \left( \frac{\nu -1}{2\nu}
\right)^{\frac{\nu}{\nu-1}}~,
\end{equation}

\vskip \baselineskip
\noindent
leading, in the case $\nu=2$, to
$B_{2}=1/15$~\cite{randradeepl2014,curadopre2014,nobrepre2015}.
Now, manipulating this result together with the internal
energy of~\eq{eq:intennu}, one may express
$s_{\nu} \equiv s_{\nu}(u,\alpha)$,

\begin{equation}
\label{eq:entu}
s_{\nu}(u,\alpha) =  \frac{k}{\nu-1}
\left\{ 1- B_{\nu} [C_{\nu}]^{\frac{\nu(\nu+1)}{\nu-1}}
\left[\frac{\alpha \lambda^{2} (\nu-1)}{2u(3\nu-1)}\right]^{\frac{\nu-1}{2}} \right\}~,
\end{equation}

\vskip \baselineskip
\noindent
where the dependence $s_{\nu} \equiv s_{\nu}(u,\alpha)$ will
become clear later on.

From~\eq{eq:entu} one obtains the fundamental thermodynamic relation,

\begin{equation}
\label{eq:theta2}
\left(\frac{\partial s_{\nu}}{\partial u}\right)_{\alpha}=\frac{1}{\theta}~,
\end{equation}

\vskip \baselineskip
\noindent
which is analogous to the temperature definition of standard
thermodynamics~\cite{reif,balian,reichl}, showing that the
parameter $\theta$ introduced in~\eq{eq:gentemp}
represents an appropriate effective-temperature definition for
any $\nu>1$.

The result of~\eq{eq:theta2} suggests a definition of a type of
energy exchange, $\delta Q=\theta ds_{\nu}$, to be called hereafter as
heat exchange, following previous
works~\cite{randradeepl2014,curadopre2014,nobrepre2015}.
In what follows, we will use the
results above to construct a thermodynamical framework for
the present system of interacting particles.

\section{Thermal Contact Between Systems and Zeroth Law}

\subsection{Systems in Thermal Contact}

In order to study the thermal contact between two systems
we consider the same physical situation introduced in Ref.~\cite{nobrepre2015},
for the case $\nu=2$. It consists in two rectangular systems
(to be called herein as systems 1 and 2),
containing $N_{1}$ and $N_{2}$ particles, respectively,
and being characterized
by equal sizes in the $x$-direction $(L_{x})$ and different sizes in the
$y$-direction, $(L_{y}^{(1)},L_{y}^{(2)})$.
The two systems are put together through a contact along the
$x$-direction; the size $L_{x}$ is kept fixed, whereas $L_{y}^{(1)}$
and $L_{y}^{(2)}$ are allowed to change by means of a movable,
rigid, and impermeable wall that separates the two systems.
Since there is no exchange of particles between the two systems,
variations in the corresponding effective temperatures of
each system may occur only due to changes in their $y$-direction sizes,
according to~\eq{eq:gentemp}; herein we refer to this apparatus as a
thermal contact between systems 1 and 2.

The physical transformation consists in a change
in the lengths in the $y$-direction, from a
state with initial lengths $(L_{y,i}^{(1)},L_{y,i}^{(2)})$,
to a state with final lengths $(L_{y,f}^{(1)},L_{y,f}^{(2)})$,

\begin{equation}
\label{eq:deltaly}
L_{y,f}^{(1)} = L_{y,i}^{(1)} - \delta L_{y}~; \qquad
L_{y,f}^{(2)} = L_{y,i}^{(2)} + \delta L_{y} \qquad (\delta L_{y} > 0)~,
\end{equation}

\vskip \baselineskip
\noindent
by conserving their sum,

\begin{equation}
\label{eq:conslenghts}
L_{y,f}^{(1)} + L_{y,f}^{(2)} = L_{y,i}^{(1)} + L_{y,i}^{(2)}~.
\end{equation}

\vskip \baselineskip
\noindent
Hence, the transformation occurs in such a way
that the system with a higher density of particles pushes
the wall, leading to a mechanical equilibrium
characterized by~\cite{nobrepre2015},

\begin{equation}
\label{eq:theta1eqtheta2}
{N_{1} \over L_{y,f}^{(1)}} = {N_{2} \over L_{y,f}^{(2)}}~,
\end{equation}

\vskip \baselineskip
\noindent
which corresponds precisely to
$\theta_{f}^{(1)} = \theta_{f}^{(2)} = \theta_{f}$.

Using the conservation of the total length in~\eq{eq:conslenghts},
together with the definition
of effective temperature in~\eq{eq:gentemp}, one may write

\begin{equation}
\label{eq:thetafn1n2}
[\theta_{f}]^{-\frac{1}{\nu-1}}
= {N_{1} \over N_{1}+N_{2}} \, [\theta_{i}^{(1)}]^{-\frac{1}{\nu-1}}
+ {N_{2} \over N_{1}+N_{2}} \, [\theta_{i}^{(2)}]^{-\frac{1}{\nu-1}}~,
\end{equation}

\vskip \baselineskip
\noindent
which may be expressed also in terms of the initial lengths
$(L_{y,i}^{(1)},L_{y,i}^{(2)})$,

\begin{equation}
\label{eq:thetafl1l2}
[\theta_{f}]^{\frac{1}{\nu-1}} =
\frac{L_{y,i}^{(1)}}{L_{y,i}^{(1)}+L_{y,i}^{(2)}} \, [\theta_{i}^{(1)}]^{\frac{1}{\nu-1}}
+\frac{L_{y,i}^{(2)}}{L_{y,i}^{(1)}+L_{y,i}^{(2)}} \, [\theta_{i}^{(2)}]^{\frac{1}{\nu-1}}~.
\end{equation}

\vskip \baselineskip
\noindent
One should call the attention to Eqs.~(\ref{eq:thetafn1n2})
and~(\ref{eq:thetafl1l2}), where the final equilibrium
temperature is related to the two initial temperatures.
In both cases one finds a structure typical of average values,
where each contribution on the right-hand-side appears
multiplied by a weight factor, corresponding to
a characteristic fraction of each system, associated
either to its number of particles [Eq.~(\ref{eq:thetafn1n2})],
or to its initial length occupied [Eq.~(\ref{eq:thetafl1l2})].
Moreover, this structure is very similar to the one that appears from
the exchange of heat between two substances $A$ and $B$
in standard thermodynamics:
considering two substances with thermal
capacities $C_{A}$ and $C_{B}$, and at initial temperatures
$T_{A}$ and $T_{B}$, respectively,
the final equilibrium temperature
$T_{f}$ attained after their thermal contact
is given by~\cite{reif},

\begin{equation}
\label{eq:tfab}
T_{f} = \frac{C_{A}}{C_{A}+C_{B}} \, T_{A}
+\frac{C_{B}}{C_{A}+C_{B}} \, T_{B}~.
\end{equation}

\vskip \baselineskip
\noindent

Both Eqs.~(\ref{eq:thetafn1n2}) and~(\ref{eq:thetafl1l2}) imply that
 for a thermal contact where both systems are
initially at the same temperature, i.e., $\theta_{i}^{(1)}=\theta_{i}^{(2)}$,
then the final temperature remains unchanged,
$\theta_{f}=\theta_{i}^{(1)}=\theta_{i}^{(2)}$. Furthermore, they
guarantee the most desirable result that the final
equilibrium temperature
$\theta_{f}$ must lie inside the interval defined by the two
initial temperatures $\theta_{i}^{(1)}$ and
$\theta_{i}^{(2)}$, for any $\nu > 1$. As an example,
lets consider Eq.~(\ref{eq:thetafl1l2}), for system 2
at an initial temperature $\theta_{i}^{(2)}>\theta_{i}^{(1)}$,
which we express by $\theta_{i}^{(2)}= l \theta_{i}^{(1)}$,
where $l>1$ is a dimensionless real number; obviously,
the case $l=1$ recovers $\theta_{f}=\theta_{i}^{(1)}=\theta_{i}^{(2)}$.
Hence, since $l>1$, Eq.~(\ref{eq:thetafl1l2}) leads to

\begin{equation}
\label{eq:thetafl1l2a}
\theta_{f} = \theta_{i}^{(1)} \left[
\frac{L_{y,i}^{(1)}+L_{y,i}^{(2)} \, l^{\frac{1}{\nu-1}}}{L_{y,i}^{(1)}+L_{y,i}^{(2)}}
\right]^{\nu-1}~,
\end{equation}

\vskip \baselineskip
\noindent
which implies that $\theta_{f}>\theta_{i}^{(1)}$. Similarly, expressing
$\theta_{i}^{(1)}= \theta_{i}^{(2)}/l$, Eq.~(\ref{eq:thetafl1l2})
may be written as

\begin{equation}
\label{eq:thetafl1l2b}
\theta_{f} = \theta_{i}^{(2)} \left[
\frac{(L_{y,i}^{(1)}/l^{\frac{1}{\nu-1}})+L_{y,i}^{(2)}}{L_{y,i}^{(1)}+L_{y,i}^{(2)}}
\right]^{\nu-1}~,
\end{equation}

\vskip \baselineskip
\noindent
yielding $\theta_{f}<\theta_{i}^{(2)}$, since $l^{\frac{1}{\nu-1}}>1$;
consequently, one concludes that
$\theta_{i}^{(1)}<\theta_{f}<\theta_{i}^{(2)}$.

Readily from Eq.~(\ref{eq:thetafn1n2}), one may define a heat reservoir
as a system with a number of particles much larger than the other
one; e.g., $N_{1} \gg N_{2}$.
This condition characterizes
system 1 as a heat reservoir, a concept
introduced qualitatively in Ref.~\cite{curadopre2014}.
In this case, one has, as a first approximation for the denominators,
$N_{1}+N_{2} \approx N_{1}$,
so that~\eq{eq:thetafn1n2} becomes

\begin{equation}
\label{eq:thetafn1n2exp1}
[\theta_{f}]^{-\frac{1}{\nu-1}}
\approx  [\theta_{i}^{(1)}]^{-\frac{1}{\nu-1}}
+ {N_{2} \over N_{1}} \, [\theta_{i}^{(2)}]^{-\frac{1}{\nu-1}}~,
\end{equation}

\vskip \baselineskip
\noindent
leading to

\begin{equation}
\label{eq:thetafn1n2exp2}
\theta_{f} \approx \theta_{i}^{(1)} \left[ 1- (\nu-1) {N_{2} \over N_{1}} \, \left( {\theta_{i}^{(2)} \over \theta_{i}^{(1)}}
\right)^{-\frac{1}{\nu-1}} \right]~,
\end{equation}

\vskip \baselineskip
\noindent
showing that $\theta_{f} = \theta_{i}^{(1)} + O(N_{2}/N_{1})$, so that
the final temperature is essentially the temperature of the reservoir.

\subsection{Zeroth Law}

Let us consider two systems (1 and 2) in contact, as
defined above, so that their mechanical equilibrium
is characterized by

\begin{equation}
\label{eq:zerothlaw12}
{N_{1} \over L_{y,1}} = {N_{2} \over L_{y,2}}~.
\end{equation}

\vskip \baselineskip
\noindent
Now, by bringing a third system (system 3) under the same
contact defined above, and in equilibrium with
system 1, one has

\begin{equation}
\label{eq:zerothlaw13}
{N_{1} \over L_{y,1}} = {N_{3} \over L_{y,3}}~.
\end{equation}

\vskip \baselineskip
\noindent
Hence, the right-hand sides of
Eqs.~(\ref{eq:zerothlaw12}) and~(\ref{eq:zerothlaw13})
are equal to one another, so that
systems 2 and 3 are also in equilibrium, i.e.,
$(N_{2}/L_{y,2})=(N_{3}/L_{y,3})$.

This result enables 1 as a test system,
i.e., a thermometer, so that all systems in equilibrium
with 1, should be in equilibrium with each other.
These systems have a property in common, which
is their effective temperature $\theta$, defined
in~\eq{eq:gentemp}. Therefore, one can now formulate
the zeroth principle, in a more general context (valid
for $\nu>1$), with
respect to the previous formulation of Ref.~\cite{nobrepre2015}
(restricted to $\nu=2$):

\vskip \baselineskip
\noindent
{\it Two systems of interacting particles, for which correlations
are taken into account and thermal effects are negligible in comparison
with those associated with their effective temperatures, are said to be in
thermal equilibrium if, being in contact, no heat flows in either way.
The zeroth principle can be enunciated by stating that two of these
systems in thermal equilibrium with a third one, are in thermal equilibrium with
each other.}

\section{First-Law Proposal}

\subsection{First Law and Equation of State}

Previous studies have proposed a work contribution, related to the
external potential, given by variations in the parameter
$\alpha$~\cite{curadopre2014,randradeepl2014,nobrepre2015}.
This suggestion was based on the fact that variations in
$\alpha$ change the cutoff value $x_{e}$ [see, e.g.,~\eq{eq:xe}], and
consequently, control directly the volume occupied by the particles in
the equilibrium state.
In this way, an infinitesimal change in $\alpha$, modifies
the external potential acting on each particle, leading to an infinitesimal work
defined as $\delta W=\sigma d\alpha$, where $\sigma$ represents
a parameter thermodynamically
conjugated to $\alpha$, to be determined by means of an equation of state.
Moreover, the fundamental relation of~\eq{eq:theta2} leads to a
definition of a type of
energy exchange, i.e., an infinitesimal heat, $\delta Q=\theta ds_{\nu}$.
These two types of energy exchange
yield a proposal equivalent to the First
Law~\cite{randradeepl2014,curadopre2014,nobrepre2015},

\begin{equation}
\label{1law}
du=\delta Q + \delta W=\theta ds_\nu + \sigma d\alpha~,
\end{equation}

\vskip \baselineskip
\noindent
where $\delta W$ corresponds to the work done \textit{on} the system,
and $\sigma$ should present units, $[\sigma]_{\cal D}=[{\rm length}]^{2}$.
In this way, the dependence $s_{\nu}=s_{\nu}(u,\alpha)$ [cf.~\eq{eq:entu}]
becomes clear; the consistency of this proposal for the first law
will be shown throughout the next sections.

Considering Eq.~(\ref{1law}) for $u$ fixed, one gets that

\begin{equation}
\label{eq:dsdalpha}
\left(\frac{\partial s_{\nu}}{\partial
\alpha}\right)_{u}=-\frac{\sigma}{\theta}~,
\end{equation}

\vskip \baselineskip
\noindent
so that deriving Eq. (\ref{eq:entu}) with respect to $\alpha$ and using the
internal energy of~\eq{eq:intennu}, one obtains,

\begin{equation}
\label{eq:eqstate}
\sigma = {\nu-1 \over 6\nu-2} \, C_{\nu}^{2}
\lambda^2 \left( \frac{k\theta }{\alpha\lambda^2} \right)^{2/(\nu+1)}~,
\end{equation}

\vskip \baselineskip
\noindent
which represents the equation of state for the present system.
Therefore, the parameter $\sigma$ is always positive for
$\nu>1$, and consequently, the work contribution
$\delta W=\sigma d\alpha$ is positive (negative) for transformations
that increase (decrease) $\alpha$. Moreover,
from~\eq{eq:eqstate} one sees that $\sigma$ increases with $\theta$
(for $\alpha$ fixed),  whereas for a fixed $\theta$,
an increase in $\sigma$ yields a decrease in $\alpha$.

Comparing Eqs.~(\ref{eq:intennu}) and~(\ref{eq:eqstate}) one finds
that curious relation,

\begin{equation}
\label{eq:relusigma}
u=\sigma \alpha~,
\end{equation}

\vskip \baselineskip
\noindent
which was found also in the case $\nu=2$, for both
harmonic~\cite{curadopre2014,nobrepre2015}
and non-harmonic~\cite{ribeiropre2016} external
potentials. This relation will have
implications on the thermodynamic potentials, particularly, on the enthalpy,
to be discussed later on.
Furthermore, Eqs.~(\ref{eq:averx2}) and~(\ref{eq:eqstate}) lead to

\begin{equation}
\label{eq:relx2sigma}
\langle x^{2} \rangle =2 \sigma~,
\end{equation}

\vskip \baselineskip
\noindent
showing that the parameter $\sigma$ acts directly on the equilibrium
distribution.

\subsection{Carnot Cycle}

\begin{figure}
\includegraphics[height=8cm]{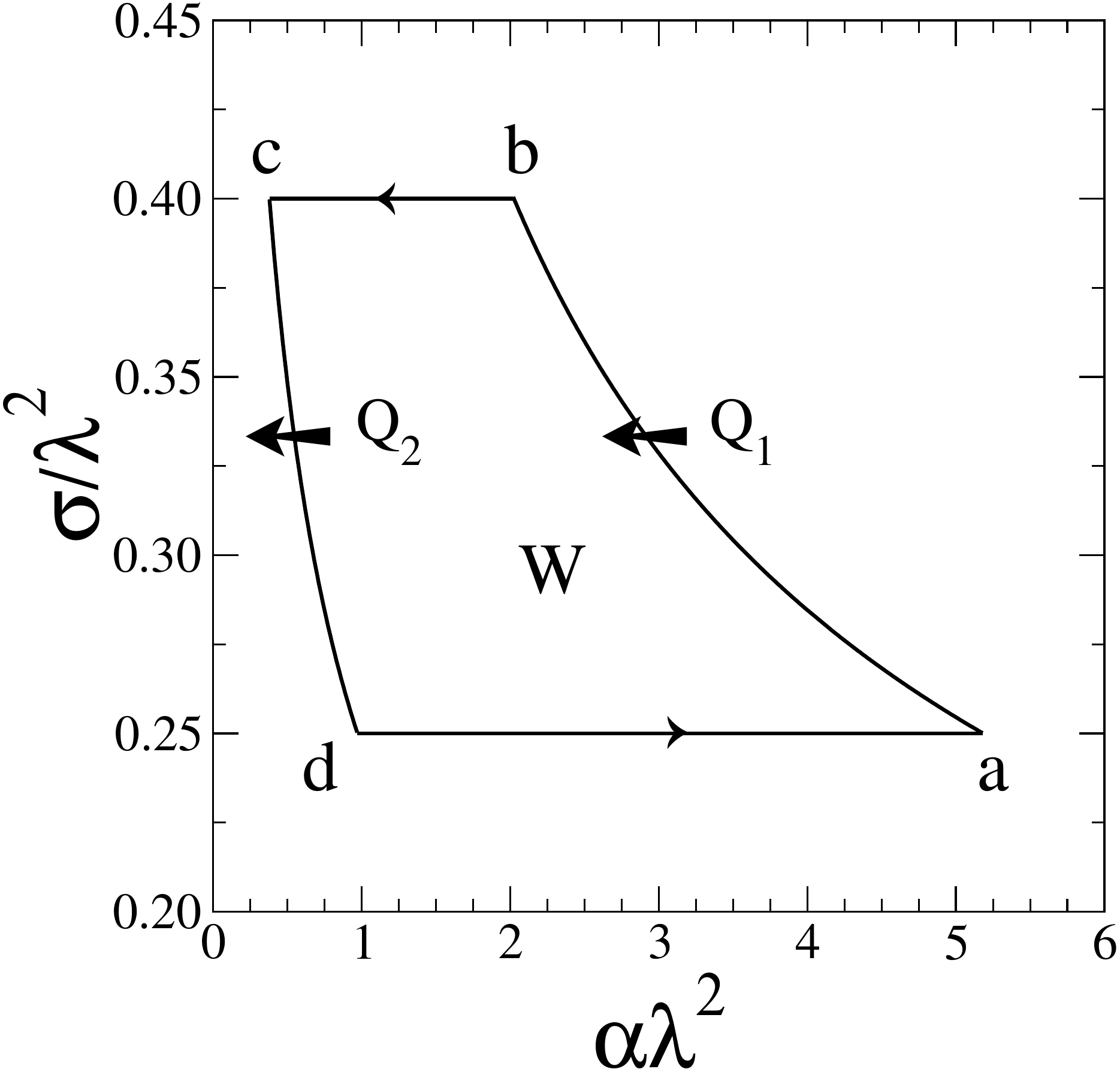}
\protect\caption{(Color online) The Carnot cycle $a \rightarrow b
\rightarrow c \rightarrow d \rightarrow a$ is represented in the
particular case $\nu=3$ (i.e., $q=-1$). The transformations for
$\sigma$ constant are adiabatic, and herein they were chosen to
occur for $(\sigma/\lambda^{2})=0.40$ ($b \rightarrow c$) and
$(\sigma/\lambda^{2})=0.25$ ($d \rightarrow a$). In general,
from~\eq{eq:eqstate} isothermal transformations correspond to
$(\sigma/\lambda^{2}) \sim (\alpha \lambda^{2})^{-2/(\nu+1)}$, and
for the particular case shown they are characterized by
$(\sigma/\lambda^{2}) \sim (\alpha \lambda^{2})^{-1/2}$. These
transformations occur at two conveniently chosen temperatures,
$\theta_{1}>\theta_{2}$, in such a way that an amount of heat
$Q_{1}$ is absorbed (released) in the isothermal process at the
higher (lower) temperature $\theta_{1}$ ($\theta_{2}$). The area
inside the cycle represents the total work $W$ done {\it on} the
system, which is negative, as expected from~\eq{1law}. The cycle
above holds for any system of units, e.g., one may consider all
quantities with dimensions of energy in Joules.}
\label{fig:carnotcycle}
\end{figure}

In order to be useful from the point of view of thermodynamics, the
entropy $s_{\nu} \equiv s_{\nu}(u,\alpha)$ should be a state
function.
Hence, standard thermodynamic manipulations
by considering the fundamental relation of~\eq{eq:theta2}, together
with~\eq{eq:dsdalpha},
and using Eqs.~(\ref{eq:eqstate}) and~(\ref{eq:relusigma}),
yield that the following second
derivatives are independent of the order of differentiation,

\begin{equation}
\label{eq:d2sdalphadu}
\frac{\partial^2 s_{\nu}}{\partial\alpha\partial u}=\frac{\partial^2 s_{\nu}}{\partial
u\partial\alpha}~.
\end{equation}

\vskip \baselineskip
\noindent
From the equation above, one obtains

\begin{equation}
\left(\frac{\partial}{\partial \alpha}\right)_{u}\left({1 \over \theta} \right)
= - \left(\frac{\partial}{\partial u}\right)_{\alpha}\left({\sigma \over \theta} \right)
={\nu -1 \over 2 \alpha \theta}~.
\end{equation}

\vskip \baselineskip
\noindent
The result of~\eq{eq:d2sdalphadu} enables the entropy $s_{\nu}$ as a
state function, so that its change due to a
given reversible thermodynamical transformation, taking the system
from an initial state $i$ to a final state $f$, depends only
on its respective values at these states, i.e.,
$\Delta s_{\nu} = s_{\nu,f} - s_{\nu,i}$.

Next, we will use the fact that the entropy $s_{\nu}$ is a state function
in order to analyze the Carnot cycle, constructed by two isothermal
$(\theta \ {\rm constant})$ and two adiabatic $(s_{\nu} \ {\rm constant})$
transformations, intercalated.
From~\eq{eq:enttheta} one sees that an adiabatic transformation
occurs for $(k\theta/\alpha \lambda^{2})=$ constant, which
corresponds to $\sigma=$ constant in the equation of state [Eq.~(\ref{eq:eqstate})];
also from this equation, isothermal transformations correspond to
$(\sigma/\lambda^{2}) \sim (\alpha \lambda^{2})^{-2/(\nu+1)}$.
Such a cycle was studied in
Refs.~\cite{randradeepl2014,curadopre2014,nobrepre2015,ribeiropre2016}
for the case $\nu=2$, where the adiabatic and isothermal transformations
were given by $\sigma=$ constant and
$(\sigma/\lambda^{2}) \sim (\alpha \lambda^{2})^{-2/3}$, respectively;
below, the Carnot cycle will be shown to hold for $\nu>1$.

In Fig.~\ref{fig:carnotcycle} we illustrate this cycle for the particular
case $\nu=3$, i.e., $q=-1$, where the adiabatic transformations correspond
to $\sigma = {\rm constant}$, whereas the two isothermal transformations
occur at temperatures $\theta_{1}$ and $\theta_{2}$, with
$\theta_{1}>\theta_{2}$.
The quantities $\sigma$ and $\alpha$, as defined above,
are such that the ordinate
$\sigma/\lambda^{2}$ is dimensionless, whereas the abscissa
$\alpha \lambda^{2}$ presents dimensions of energy.
Although shown in Fig.~\ref{fig:carnotcycle} for $\nu=3$, the
following properties hold for any $\nu>1$.
(i) An amount of heat $Q_{1}$ is absorbed in the
isothermal process at the higher temperature $\theta_{1}$, whereas the system
releases heat $Q_{2}$ in the isothermal process at the lower temperature
$\theta_{2}$. (ii) In a plot $\sigma$ versus $\alpha$ [or equivalently,
$\sigma/\lambda^{2}$ versus $\alpha \lambda^{2}$, as
in Fig.~\ref{fig:carnotcycle}],
the work associated with a given process corresponds to the area
below such transformation.
Since $\sigma$ is always positive for $\nu>1$ [cf.~\eq{eq:eqstate}],
work is positive (negative) for transformations
that increase (decrease) $\alpha$.
Therefore, the total work done {\it on} the
system, calculated as $W=W_{ab}+W_{bc}+W_{cd}+W_{da}$, is
given by the area
enclosed in the cycle of Fig.~\ref{fig:carnotcycle}, being negative.
If one defines ${\cal W}=-W$ as the work done {\it by}
the system, considering that the variation of internal energy is zero
for the complete
cycle, one has $Q_{1}={\cal W}+Q_{2}$ (conventionalizing  all these three
quantities as positive).
(iii) Hence, in the two isothermal processes, one has that

\begin{equation}
\label{eq:q1q2}
{Q_{1} \over Q_{2}}
= {\theta_{1} |\Delta s_{\nu,1}| \over \theta_{2} |\Delta s_{\nu,2}|}~,
\end{equation}

\vskip \baselineskip
\noindent
where $\Delta s_{\nu,1}$ and $\Delta s_{\nu,2}$ represent
the entropy variations in the isothermal processes
1 and 2, respectively. Since the entropy $s_{\nu}$ is a
state function, its total variation in the complete cycle should
be zero, so that $\Delta s_{\nu,1}=-\Delta s_{\nu,2}$.
Consequently, one has
$(Q_{1} / Q_{2})=(\theta_{1} / \theta_{2})$, leading to
the celebrated efficiency of the Carnot cycle,

\begin{equation}
\label{eq:effcarnot}
\eta= {{\cal W} \over Q_{1}}={Q_{1}-Q_{2} \over Q_{1}} =
1-{\theta_{2} \over \theta_{1}}~; \qquad (0 \leq \eta \leq 1)~,
\end{equation}

\vskip \baselineskip
\noindent
which holds for any $\nu>1$.
This reinforces the fact that the Carnot cycle is very special
within thermodynamics, so
that its efficiency does not depend on the system under study,
external potential, or particular entropic form considered.
As shown in Refs.~\cite{curadopre2014,nobrepre2015},
by following the cycle in its reverse way,
the corresponding Carnot refrigerator is obtained;
one may carry the same procedure of these
previous works to extend the validity of the Carnot refrigerator
for any $\nu>1$.
These results give further support for the fundamental
relation of Eq.~(\ref{1law}), as well as for the effective-temperature definition
of~\eq{eq:gentemp}, as appropriate for a consistent thermodynamical framework
of the present system.

\subsection{Efficiency of a Carnot Cycle for a General Entropic Form}

Next, we will show that the efficiency of~\eq{eq:effcarnot} may apply to a
system described by a  general entropic form
$s_{\{ \mu \}}[P]$, where $\{ \mu \}$ stands for a given set of real indexes that
defines such an entropy, whereas $[P]$ represents either
a discrete set of probabilities, or
a continuous probability distribution (for a detailed classification of
entropic forms see, e.g., Ref.~\cite{thurnerepl2011a}).
Let us then consider such a system under the presence of external parameters
that can be varied in order to produce work on it; we will restrict ourselves to
the simplest case of a single external parameter, to be denoted
by $\alpha$, similarly to the example above.
Herein, we presuppose that the equilibrium entropy is a state function, and that it
may be expressed in the form
$s_{\{ \mu \}} \equiv s_{\{ \mu \}}(u,\alpha)$ [like the one in~\eq{eq:entu}],
 so that the fundamental relation of~\eq{eq:theta2} follows,
leading to a definition of an effective temperature $\theta$.

In this way, we now consider a Carnot cycle, constructed by two
isothermal ($\theta={\rm constant}$) and two adiabatic
[$s_{\{ \mu \}}(u,\alpha)={\rm constant}$] transformations, intercalated.
The isothermal transformations occur for effective temperatures
$\theta_{1}$ and $\theta_{2}$, with $\theta_{1}>\theta_{2}$.
The system under investigation should be able to exchange energy with
its environment, by transferring heat $\delta Q$, and performing
work $\delta W$, according to an infinitesimal form of the
first law, like Eq.~(\ref{1law}).
The infinitesimal heat is given by $\delta Q=\theta ds_{\{ \mu \}}$, whereas
the infinitesimal quantity of work may be defined as $\delta W=\sigma d\alpha$,
with $\sigma$ representing the
parameter thermodynamically conjugated to $\alpha$.
Since the entropy is a state function, its change along the whole
cycle is zero, leading to
$\Delta s_{\{ \mu \},1}=-\Delta s_{\{ \mu \},2}$,
where 1 and 2 refer to the two isothermal processes of the cycle.
Similarly to what was done in Fig.~\ref{fig:carnotcycle}, by
considering that the variation of internal energy is zero for the complete
cycle, one uses Eq.~(\ref{1law}) to obtain
$Q_{1}={\cal W}+Q_{2}$ (conventionalizing  all these three
quantities as positive), and using the relation of~\eq{eq:theta2}
for the two isothermal processes,
$Q_{1} = \theta_{1} |\Delta s_{\{ \mu \},1}|$
and $Q_{2} = \theta_{2} |\Delta s_{\{ \mu \},2}|$, respectively.
From these results, Eq.~(\ref{eq:q1q2}) follows, leading to
the efficiency of the associated Carnot cycle in Eq.~(\ref{eq:effcarnot}).

Hence, we have shown that the well-known expression for the
efficiency of a Carnot cycle
applies to a general entropic form $s_{\{ \mu \}}[P]$;
such a demonstration is expected to be applicable to a cycle composed by a
working substance within the realm of
complex systems, which have been usually associated to these types of
entropic forms.

In what follows we continue with the thermodynamic framework
associated with the entropy
of~\eq{eq:tsallisentnu} and the corresponding equilibrium
distribution of~\eq{eq:tsallisdist}.

\section{Thermodynamic Potentials and  Response Functions}

Now we will explore the first-law proposal of Eq.~(\ref{1law}), by
following usual procedures~\cite{reichl,balian,reif}, e.g.,
performing Legendre transformations, in order to introduce further
thermodynamic potentials. Additionally, we
define quantities analogous to the response functions of standard
thermodynamics.

\subsection{Internal energy}

From Eq.~(\ref{1law}) one has that the internal energy depends on the pair of
independent variables $(s_\nu,\alpha)$, i.e., $u \equiv u(s_\nu,\alpha)$;
indeed, inverting Eq.~(\ref{eq:entu}) one gets,

\begin{equation}
\label{eq:intenent}
u(s_{\nu},\alpha) =   \left\{ B_{\nu}
[C_{\nu}]^{\frac{\nu(\nu+1)}{\nu-1}} \right\}^{\frac{2}{\nu-1}}
\frac{(\nu-1)\alpha\lambda^2}{2(3\nu-1)[1-(\nu-1)s_{\nu}/k]^{\frac{2}{\nu-1}}}~.
\end{equation}

\vskip \baselineskip
\noindent
From the expression above one obtains equations that are equivalent
to the effective-temperature definition [Eq.~(\ref{eq:theta2})], as well as
the equation of state [Eq.~(\ref{eq:eqstate})],
respectively,

\begin{equation}
\left(\frac{\partial u}{\partial
s_\nu}\right)_\alpha=\theta~; \qquad \left(\frac{\partial
u}{\partial\alpha}\right)_{s_\nu} =\sigma~.
\end{equation}

\vskip \baselineskip
\noindent
The internal energy $u(s_\nu,\alpha)$ must be a state function;
for that, its second derivatives
should be independent of the order of differentiation, leading to the
following Maxwell relation,

\begin{equation}
\frac{\partial^2 u}{\partial\alpha\partial s_\nu}=\frac{\partial^2 u}{\partial
s_\nu\partial\alpha} \quad \Rightarrow \quad \left(\frac{\partial\sigma}{\partial
s_\nu}\right)_\alpha=\left(\frac{\partial\theta}{\partial\alpha}\right)_{s_\nu}~.
\end{equation}

\subsection{Helmholtz free energy}

Now we consider the Helmholtz free energy, $f(\theta,\alpha)$ [defined
in~\eq{eq:freeenergy}],

\begin{equation}
f(\theta,\alpha)=u-\theta s_\nu \quad\Rightarrow \quad
df=-s_\nu d\theta+\sigma d\alpha~.
\end{equation}

\vskip \baselineskip
\noindent
The following free energy results,

\begin{equation}
\label{eq:helmfreeen}
f_{\nu}(\theta,\alpha) =   \left\{
\frac{(\nu-1)[C_{\nu}]^{2}}{2(3\nu-1)} +
\frac{B_{\nu}[C_{\nu}]^{\frac{3\nu-1}{\nu-1}}}{\nu-1}  \right\}
\left( \alpha \lambda^{2} \right)^{\frac{\nu-1}{\nu+1}} \left( k
\theta \right)^{\frac{2}{\nu+1}} -\frac{k\theta}{\nu-1}~,
\end{equation}

\vskip \baselineskip
\noindent
which satisfies the relations,

\begin{equation}
\left(\frac{\partial
f}{\partial\theta}\right)_\alpha=-s_\nu~; \qquad \left(\frac{\partial
f}{\partial\alpha}\right)_{\theta} =\sigma~.
\end{equation}

\vskip \baselineskip
\noindent
Furthermore, the corresponding Maxwell relation appears,

\begin{equation}
\frac{\partial^2 f}{\partial\alpha\partial\theta}=\frac{\partial^2
f}{\partial\theta\partial\alpha} \quad \Rightarrow \quad
\left(\frac{\partial s_\nu}{\partial
\alpha}\right)_\theta=-\left(\frac{\partial\sigma}{\partial\theta}\right)_{
\alpha}~.
\end{equation}

\subsection{Gibbs free energy}

We define the Gibbs potential $g(\theta,\sigma)$ through

\begin{equation}
g(\theta,\sigma)=f-\sigma\alpha=u-\theta s_\nu-\sigma\alpha
\quad \Rightarrow \quad
dg=-s_\nu d\theta-\alpha d\sigma~.
\end{equation}

\vskip \baselineskip
\noindent
In fact, using the relation of~\eq{eq:relusigma}, one has that
$g=-\theta s_{\nu}$, so that~\eq{eq:entu} yields

\begin{equation}
\label{eq:gibbspotential}
g(\theta,\sigma) = \frac{k \theta}{\nu-1} \left\{
B_{\nu} [C_{\nu}]^{\frac{\nu(\nu+1)}{\nu-1}} \left[ \frac{
\lambda^{2} (\nu-1)}{2\sigma(3\nu-1)} \right]^{\frac{\nu-1}{2}} - 1
\right\}~,
\end{equation}

\vskip \baselineskip
\noindent
which satisfies

\begin{equation}
\left(\frac{\partial
g}{\partial\theta}\right)_\sigma=-s_\nu~; \qquad \left(\frac{\partial
g}{\partial\sigma}\right)_{\theta} =-\alpha~.
\end{equation}

\vskip \baselineskip
\noindent
The first of these equations follows trivially from $g=-\theta s_{\nu}$,
whereas the second leads to the equation
of state of Eq.~(\ref{eq:eqstate}).
The Maxwell relation associated with this pair of variables is
obtained as

\begin{equation}
\label{mrgibbs}
\frac{\partial^2 g}{\partial\sigma\partial\theta}=\frac{\partial^2
g}{\partial\theta\partial\sigma} \quad \Rightarrow \quad \left(\frac{\partial s_\nu}{\partial
\sigma}\right)_\theta=\left(\frac{\partial\alpha}{\partial\theta}\right)_{\sigma}~.
\end{equation}

\vskip \baselineskip
\noindent
The consequences of the relation $g=-\theta s_{\nu}$
will be discussed next.

\subsection{Enthalpy}

The relation of~\eq{eq:relusigma}, i.e., $u=\sigma\alpha$,
has led to a Gibbs potential $g=-\theta s_{\nu}$;
these relations imply on a trivial enthalpy, defined as

\begin{equation}
h(s_\nu,\sigma)=u-\sigma\alpha=f+\theta s_\nu-\sigma\alpha
= g+\theta s_\nu = 0~.
\end{equation}

\vskip \baselineskip
\noindent
Hence,

\begin{equation}
\label{eq:dsdsigma}
dh=\theta ds_\nu-\alpha d\sigma = 0 \quad \Rightarrow \quad
ds_\nu = {\alpha \over \theta} \; d\sigma~,
\end{equation}

\vskip \baselineskip
\noindent
showing that variations in $\sigma$ are directly related to variations in
the entropy $s_{\nu}$, reinforcing the previous result that
for an adiabatic process
the present system cannot exchange ``heat'' (i.e., it cannot
vary its entropy) for $\sigma$ fixed.

Therefore, a complete thermodynamic framework is given in terms of
the three previously-defined potentials, namely,
internal energy, $u(s_{\nu},\alpha)$, and free energies,
$f(\theta,\alpha)$, and $g(\theta,\sigma)$. As already verified in
Ref.~\cite{nobrepre2015}
for the case $\nu=2$, the enthalpy
$h(s_{\nu},\sigma)$ is trivial
for any $\nu>1$ and does not contain any new information
for the present system.
One should remind that a similar behavior is also found
in other well-known systems, like in the three-dimensional
ideal gas, for which
$pv=2u/3$ (valid for the classic case, as well as in
both quantum statistics~\cite{balian,reichl}), leading
to an enthalpy $h=5u/3$, showing that for an ideal gas,
the enthalpy does not represent an independent
thermodynamic potential.

\subsection{Response functions}

A quantity analogous to the specific heat was introduced in
Refs.~\cite{nobrepre2012,curadopre2014,nobrepre2015} for
a fixed $\alpha$; it may be calculated in three different ways,

\begin{equation}
\label{eq:spheatalpha}
c_{\alpha} = \left( \frac{\partial
u}{\partial \theta} \right)_{\alpha} =\theta\left(\frac{\partial s_{\nu}}{\partial
\theta}\right)_\alpha=-\theta\left(\frac{\partial^2
f}{\partial^2\theta}\right)_\alpha=\frac{k(\nu-1)}{(\nu+1)(3\nu-1)} \,
C_{\nu}^{2} \left( \frac{\alpha\lambda^2}{ k\theta } \right)^{
\frac{\nu-1}{\nu+1} } ,
\end{equation}

\vskip \baselineskip
\noindent
being a positive quantity for $\nu>1$. In a similar way, one can
define $c_\sigma$,

\begin{equation}
\label{csigma}
c_\sigma=\theta\left(\frac{\partial s_{\nu}}{\partial
\theta}\right)_\sigma=\left(\frac{\partial h}{\partial\theta}\right)_\sigma=0~,
\end{equation}

\vskip \baselineskip
\noindent
following trivially as a consequence of the fact that $h=0$, and expected, since
the system cannot exchange heat for $\sigma$ fixed.

Using the equation of state in~\eq{eq:eqstate} one can calculate
other response functions, corresponding
to the coefficient of expansion and isothermal compressibility
of standard thermodynamics; these quantities can be expressed
respectively, as

\begin{equation}
\label{coeffexpansion}
\gamma=\frac{1}{\alpha}\left(\frac{\partial\alpha}{\partial
\theta}\right)_\sigma = { 1\over \theta}~,
\end{equation}

\vskip \baselineskip
\noindent
and

\begin{equation}
\label{isotcompress}
\kappa=-\frac{1}{\alpha}\left(\frac{\partial
\alpha}{\partial\sigma}\right)_\theta = {\nu+1 \over 2 \sigma}~,
\end{equation}

\vskip \baselineskip
\noindent
being both positive quantities.
Moreover, one can also show that the response functions
defined above are related by

\begin{equation}
\label{calpha3}
c_\alpha=\alpha\theta \; \frac{\gamma^2}{\kappa}~.
\end{equation}

\vskip \baselineskip
\noindent
Therefore, these quantities behave very similarly to the
corresponding ones of standard thermodynamics~\cite{reif}, including the
relation between them [Eq.~(\ref{calpha3})]; these
results, shown to be valid for $\nu=2$ in Ref.~\cite{nobrepre2015},
were extended herein for any $\nu>1$.

\section{Conclusions}

Recent works have shown that systems of particles, interacting
repulsively through short-range forces
and evolving under overdamped motion, may be associated
to a nonlinear Fokker-Planck
equation within the class of nonextensive statistical mechanics.
The identification with this equation, characterized by a nonlinear
diffusion contribution whose exponent is given by $\nu=2-q$, was reached
through both a coarse-graining approach and molecular-dynamics simulations
of the equations of motion.
These procedures were applied to a system of interacting
vortices relevant for type-II superconductors, where the
exponent $\nu=2$ was
obtained~\cite{zapperiprl2001,andradeprl2010,ribeiropre2012,%
ribeiroepjb2012}, and more recently,
by introducing effects of correlations, further physical systems
corresponding to $\nu>2$ were identified in Ref.~\cite{vieirapre2016}.

Motivated by these types of physical systems, we have developed
a consistent thermodynamic framework
valid for the whole range
$\nu >1$ (i.e., $q<1$), which includes the values of the investigation
of Ref.~\cite{vieirapre2016}, as well as the particular case $\nu=2$.
In this approach, one basic requirement is an equilibrium distribution
$P_{\rm eq}(x)$ characterized by a cutoff value $x_{e}$,
which is typical of nonextensive statistical mechanics for $q<1$,
extending the framework built for
$\nu=2$~\cite{nobrepre2012,curadopre2014,%
randradeepl2014,nobrepre2015,ribeiropre2015,ribeiropre2016}.

The procedure is based on the definition of an effective
temperature $\theta$, conjugated to the entropic form $s_{\nu}$,
typical of nonextensive statistical mechanics.
Similarly to the previous works, the following framework holds
for typical effective temperatures $\theta \gg T$, so that
the usual thermal effects (associated with the temperature $T$)
can be neglected.
From the corresponding nonlinear Fokker-Planck equation,
the equilibrium distribution $P_{\rm eq}(x)$ was obtained,
which corresponds to a $q$-Gaussian distribution with $q<1$,
and it is shown that the cutoff $x_{e}$ depends on $\theta$,
behaving like $x_{e} \sim \theta^{1/(\nu+1)}$;
as a consequence, the variance
increases as $\langle x^2 \rangle \propto \theta^{2/(\nu+1)}$.
From this equilibrium distribution we have expressed thermodynamic
quantities, like entropy and internal energy, in terms of
the effective temperature $\theta$.
A similar framework associated to $q$-Gaussian distributions
with $q \geq 1$ is still missing; the appropriate finite quantity
to be related with an effective temperature $\theta$
concerns one of the main questions, particularly in those cases
where the variance diverges.

The consistency of our definitions is shown by introducing a thermal
contact between two systems, in such a way to obtain a formulation
for the Zeroth Law of thermodynamics. Moreover,
by proposing an infinitesimal form for the First Law,
we have considered Legendre transformations in order to
derive thermodynamic potentials, Maxwell relations,
obtaining well-known structures for the thermodynamic potentials.
Furthermore, we have defined physical
transformations and explored the Carnot cycle, calculating the celebrated
expression for its efficiency, $\eta=1-(\theta_2/\theta_1)$,
where $\theta_1$ and $\theta_2$ correspond to the effective temperatures
associated with two isothermal transformations,
with $\theta_1>\theta_2$.
We have shown that such a result for the efficiency of the Carnot cycle
holds for any entropic form that satisfies basic thermodynamic requirements.

Previous works have shown that a system of interacting vortices,
commonly used for modelling type-II superconductors,
is associated to an entropy $s_{2}$,
representing an important physical application for nonextensive
statistical mechanics.
The present extended thermodynamic framework,
for a system of interacting particles under the above-mentioned
conditions, and associated to an entropy $s_{\nu}$, with $\nu>1$,
certainly enlarges the possibility of experimental verifications.
As potential candidates, one could mention complex physical systems,
like dusty plasma~\cite{shukla,goree,sheridan,morfill} and
polymer solutions~\cite{teraoka}.

\begin{acknowledgments}
The authors thank C. Tsallis for fruitful
conversations. Partial financial support from CNPq,
CAPES, and FAPERJ (Brazilian funding agencies) is acknowledged.
One of us (A.M.C.S.) also acknowledges financial support by the John Templeton Foundation (USA).
\end{acknowledgments}

\vskip 2\baselineskip

\end{document}